\documentclass[letter]{aa} 
\usepackage{graphics}
\usepackage{graphicx}
\usepackage{multicol}
\usepackage{longtable}
\usepackage{supertabular}
\usepackage{txfonts}
\usepackage{color,soul}

\usepackage{natbib,twoopt}
\usepackage{amsmath} 
\usepackage[breaklinks=true]{hyperref} 
\bibpunct{(}{)}{;}{a}{}{,} 
\makeatletter
\newcommandtwoopt{\citeads}[3][][]{\href{http://adsabs.harvard.edu/abs/#3}%
{\def\hyper@linkstart##1##2{}%
\let\hyper@linkend\@empty\citealp[#1][#2]{#3}}}
\newcommandtwoopt{\citepads}[3][][]{\href{http://adsabs.harvard.edu/abs/#3}%
{\def\hyper@linkstart##1##2{}%
\let\hyper@linkend\@empty\citep[#1][#2]{#3}}}
\newcommandtwoopt{\citetads}[3][][]{\href{http://adsabs.harvard.edu/abs/#3}%
{\def\hyper@linkstart##1##2{}%
\let\hyper@linkend\@empty\citet[#1][#2]{#3}}}
\newcommandtwoopt{\citeyearads}[3][][]%
{\href{http://adsabs.harvard.edu/abs/#3}
{\def\hyper@linkstart##1##2{}%
\let\hyper@linkend\@empty\citeyear[#1][#2]{#3}}}
\makeatother

\usepackage{color}
\definecolor{mygreen}{RGB}{0,128,0}

\hypersetup{colorlinks=true,linkcolor=blue,citecolor=blue,urlcolor=blue}

\begin{document} 

\title{VLTI/GRAVITY upper limit on near-infrared emission from the nearby $33\,M_\odot$ black hole Gaia BH3\thanks{The GRAVITY calibrated data files are available in electronic form at the CDS via anonymous ftp to \url{cdsarc.u-strasbg.fr} (130.79.128.5) or via \url{http://cdsweb.u-strasbg.fr/cgi-bin/qcat?J/A+A/}.}}

\author{
Pierre~Kervella\inst{1,4}
\and
Pasquale~Panuzzo\inst{2,1}
\and
Alexandre~Gallenne\inst{3,4}
\and
Antoine~Mérand\inst{5}
\and
Frédéric~Arenou\inst{2,1}
\and
Elisabetta~Caffau\inst{1}
\and
Sylvestre~Lacour\inst{1}
\and
Tsevi~Mazeh\inst{6}
\and
Berry~Holl\inst{7}
\and
Carine~Babusiaux\inst{8}
\and
Nicolas~Nardetto\inst{9}
\and
Maïca~Clavel\inst{8}
\and
Jean-Baptiste~Le~Bouquin\inst{8}
\and
Damien~S\'egransan\inst{7}
}

\institute{
LIRA, Observatoire de Paris, Université PSL, Sorbonne Université, Université Paris Cité, CY Cergy Paris Université, CNRS, 92190 Meudon, France.
\email{pierre.kervella@observatoiredeparis.psl.eu}.
\and
UNIDIA, Observatoire de Paris, Université PSL, CNRS, 92190 Meudon, France.\\
\email{pasquale.panuzzo@observatoiredeparis.psl.eu}.
\and
Universidad de Concepción, Departamento de Astronomía, Casilla 160-C, Concepción, Chile.
\and
French-Chilean Laboratory for Astronomy, IRL 3386, CNRS and U. de Chile, Casilla 36-D, Santiago, Chile.
\and
European Southern Observatory, Karl-Schwarzschild-Str. 2, 85748, Garching bei München, Germany.
\and
School of Physics and Astronomy, Tel Aviv University, Tel Aviv 6997801, Israel.
\and
Department of Astronomy, University of Geneva, Chemin Pegasi 51, 1290 Versoix, Switzerland.
\and
Université Grenoble Alpes, CNRS, IPAG, 38000 Grenoble, France.
\and
Université Côte d’Azur, Observatoire de la Côte d’Azur, CNRS, Laboratoire Lagrange, CS 34229, 06304 Nice Cedex 4, France.
}

\titlerunning{VLTI/GRAVITY observations of Gaia BH3}
\authorrunning{P. Kervella et al.}

\date{Received ; Accepted 6 February 2025.}

  \abstract
   {The recent astrometric discovery of the nearby (590 pc) massive ($33\,M_\odot$) dormant black hole candidate Gaia BH3 offers the possibility to angularly resolve the black hole from its companion star by using optical interferometry.}
   {Our aim is to detect emission in the near-infrared $K$ band from the close-in environment of Gaia BH3 caused by accretion.}
   {Gaia BH3 was observed with the GRAVITY instrument using the four 8-meter Unit Telescopes of the VLT Interferometer. We searched for the signature of emission from the black hole in the interferometric data using the CANDID, PMOIRED, and exoGravity tools.}
   {With a present separation of 18\,mas, the Gaia BH3 system can be well resolved angularly by GRAVITY.
   We did not detect emission from the black hole at a contrast level of $\Delta m = 6.8$\,mag with respect to the companion star, that is, $f_\mathrm{BH}/f_\star < 0.2\%$. This corresponds to an upper limit on the continuum flux density of $f_\mathrm{BH} < 1.9 \times 10^{-16}$\,W\,m$^{-2}$\,$\mu$m$^{-1}$ in the $K$ band. In addition, we did not detect emission from the black hole in the hydrogen Br$\gamma$ line.}
   {The non-detection of near-infrared emission from the black hole in Gaia BH3 indicates that its accretion of the giant star wind is presently occurring at most at a very low rate. This is consistent with the limit of $f_\mathrm{Edd}< 4.9 \times 10^{-7}$ derived previously on the Eddington ratio for an advection-dominated accretion flow. Deeper observations with GRAVITY may be able to detect the black hole as the companion star approaches periastron around 2030.}

   \keywords{Stars: individual: LS II +14 13; Stars: black holes; Techniques: interferometric; Stars: binaries; Stars: Population II}

\maketitle


\section{Introduction}

The \object{Gaia BH3} system recently discovered from pre-release \textit{Gaia} DR4 astrometry \citepads{2024A&A...686L...2G} is composed of a low-mass giant star ($0.76 \pm 0.05\,M_\odot$) and a high-mass dark component ($32.70 \pm 0.82\,M_\odot$) that is most plausibly a black hole (BH).
\textit{Gaia} repeatedly measured the position of the photocenter of Gaia BH3 (\object{LS II +14 13}, \object{Gaia DR3 4318465066420528000}, \object{2MASS J19391872+1455542}), that is, the position of the star assuming that the BH flux is negligible.
This displacement effectively traces the orbital motion of the star around the center of mass of the system (see, e.g., \citeads{2024PASP..136g3001B}).
Among the 1.5 million examined orbital solutions in the \textit{Gaia} DR4 preliminary NSS data, Gaia BH3 has the largest astrometric mass function $f_M=32.06 \pm 0.64$, indicating that this is a very exceptional object within several kiloparsecs from the Sun.
The high mass of the dark component makes it the most massive of all known BHs in the Milky Way, apart from the supermassive \object{Sagitarius A*}.
The age of the very low metallicity ($\mathrm{[Fe/H]} = -2.56 \pm 0.11$) giant star component is estimated to be $\approx 12$\,Ga\footnote{We abbreviate the word "year" as "a" from the Latin "annum" following the recommendations of IAU \citepads{1990IAUTB..20S....W}.} \citepads{2024A&A...686L...2G}.
At a distance of only 590\,pc, it is the second nearest BH after \object{Gaia BH1} (480\,pc; \citeads{2023MNRAS.518.1057E, 2023AJ....166....6C}), which has a mass of $\approx 10\,M_\odot$.
The discovery of Gaia BH3 offers an opportunity to search for emission from the close environment of this BH.
Gaia BH3 is photometrically quiet, but the BH may be accreting a fraction of the stellar wind emitted by the giant companion as observed in symbiotic binaries \citepads{2020NewAR..9101547L}, potentially resulting in measurable emission.
Remarkably, Gaia BH3 is similar to the BH merger progenitors detected through gravitational waves (GWs) that exhibit masses of 20--40$\,M_\odot$.
In particular, it is an analog of the individual $36 \pm 4$ and $31 \pm 4\,M_\odot$ BHs that produced the first GW-detected merger, GW150914 \citepads{2016PhRvL.116x1103A}.
The high mass of Gaia BH3 also fits within the expected range for the early-Universe seeds of supermassive BHs \citepads{10.1093/mnras/stae2732}. 

We report on the first interferometric observations of Gaia BH3 collected with VLTI/GRAVITY.
The observational setup and observing log are described in Sect.~\ref{obs}, and the search for an observational signature from the BH component is discussed in Sect.~\ref{analysis}. Our conclusions are presented in Sect.~\ref{conclusion}.

\section{Observations \label{obs}}

We observed Gaia BH3 on 3 June, 23 June, and 18 July 2024 using the GRAVITY beam combiner \citepads{2017A&A...602A..94G} of ESO's Very Large Telescope Interferometer (VLTI; \citeads{2014SPIE.9146E..0JM, 2018SPIE10701E..03W, 2020SPIE11446E..06H, 2022SPIE12183E..06H}) at Cerro Paranal observatory (Chile).
Although the apparent magnitude of the stellar component in Gaia BH3 is relatively bright in the infrared ($m_K = 9.08$), we employed the four 8-meter Unit Telescopes (UTs) to reach a high S/N ratio and a correspondingly high sensitivity in terms of companion contrast.
The UTs were equipped with the first generation MACAO adaptive optics system \citepads{2003SPIE.4839..174A}.
The calibrator for the interferometric transfer function \object{BD+08 4132} (\object{Gaia DR3 4296148519431266176}, \object{2MASS J19322054+0848235}) was selected using the JMMC\footnote{\url{https://www.jmmc.fr}} \texttt{SearchCal} tool \citepads{2011A&A...535A..53B, 2016A&A...589A.112C} considering its proximity to Gaia BH3 ($6.4^\circ$) and its similar magnitude in the $K$ band. BD+08 4132 was observed immediately after Gaia BH3 using an identical spectral resolution and polarization setting of GRAVITY.
The photospheres of both the giant star component in Gaia BH3 and the calibrator are unresolved by the interferometer (fringe visibility $V>99.8\%$).
A summary of the properties of Gaia BH3 and BD+08 4132 is given in Table~\ref{table:sources}, and the log of the collected observations of Gaia BH3 is presented in Table~\ref{table:obslog}.
A description of the GRAVITY data sets at each observing epoch is provided in App.~\ref{GRAVITYplots}.
In the following analysis, we consider only the data sets collected on 23 June and 18 July 2024.

\begin{table}
\caption{Summary of the properties of the giant star in LS II +14  13 (Gaia BH3) and the interferometric calibrator BD+08 4132.
\label{table:sources}}
\centering
\begin{tabular}{lccc}
\hline \hline
\noalign{\smallskip}
Source & \object{LS II +14  13} & \object{BD+08 4132} & Ref. \\
\hline
\noalign{\smallskip}
$\alpha$ (ICRS) & 19:39:18.7115 & 19:32:20.5577 & 1 \\
$\delta$ (ICRS) & +14:55:54.011 & +08:48:23.731 & 1 \\
$\varpi$ (mas) & $1.644 \pm 0.069$ & $1.179 \pm 0.015$ & 1 \\
$G$ mag. & $11.2311 \pm 0.0028$ & $10.3592 \pm 0.0028$ & 1 \\
$K_s$ mag. & $9.075 \pm 0.015$ & $7.965 \pm 0.020$ & 2 \\
$T_\mathrm{eff}$ (K) & $5212 \pm 80$ & $\approx 4800$ & 3, 4 \\
$\log g$ & $2.929 \pm 0.003$ & $\approx 2.26$ & 3, 4 \\
$\theta_\mathrm{LD}$ ($\mu$as) & $38.54 \pm 0.12$ & $126 \pm 3$ & 3, 4 \\
$\theta_\mathrm{UD,K}$ ($\mu$as) & $36.78 \pm 0.12$ & $122 \pm 3$ & 3, 4, 5 \\ 
\hline
\end{tabular}
\tablefoot{$\theta_\mathrm{LD}$ and $\theta_\mathrm{UD}$ are the limb darkened and uniform disk angular diameters, respectively.}
\tablebib{
(1)~\citetads{GaiaEDR3content};
(2)~\citetads{2006AJ....131.1163S};
(3)~\citetads{2024A&A...686L...2G};
(4)~JMMC SearchCal \citepads{2011A&A...535A..53B, 2016A&A...589A.112C};
(5)~\citetads{2011A&A...529A..75C}.}
\end{table}

\begin{table*}
\caption{Log of the observations of Gaia BH3 (LS II +14   13) and the interferometric calibrator (BD+08 4132) with GRAVITY. 
\label{table:obslog}}
\centering
\begin{tabular}{cccccclccc}
\hline \hline
\noalign{\smallskip}
 Starting UT date & MJD & Target & Cal/ & Obj/ & Exp.  & Instrumental setup & AM & $\sigma$ & $\tau_0$ \\
 & $-60000$ & & Sci & Sky & (s) & (Res, Pol, Mode) & & $(\arcsec)$ & (ms) \\
\hline
\noalign{\smallskip}
2024-06-03T08:09:17 & 464.3400 & LS II +14   13 & Sci & (OSOO) & 389 & Med, Split, Single &  1.32 & 0.51 & 3.5 \\
2024-06-03T08:49:05 & 464.3678 & BD+08 4132 & Cal & (SO) & 389 & Med, Split, Single & 1.31 & 0.29 & 2.6 \\
\noalign{\smallskip}
2024-06-23T07:33:20 & 484.3150 & LS II +14   13 & Sci & OSOO & 376 & Med, Comb, Single & 1.41 & 0.54 & 2.3 \\
2024-06-23T08:10:10 & 484.3406 & BD+08 4132 & Cal & SO & 376 & Med, Comb, Single & 1.41 & 0.73 & 2.3 \\
2024-06-23T08:30:14 & 484.3545 & LS II +14   13 & Sci & (O)SOOO & 414 & High, Comb., Single & 1.61 & 0.58 & 2.3 \\
2024-06-23T09:14:18 & 484.3851 & BD+08 4132 & Cal & OS & 414 & High, Comb, Single & 1.81 & 0.78 & 2.3 \\
\noalign{\smallskip}
2024-07-18T06:16:05 & 509.2614 & LS II +14   13 & Sci & OSOO & 376 & Med, Comb, Dual & 1.45 & 0.32 & 7.5 \\
2024-07-18T06:52:52 & 509.2868 & BD+08 4132 & Cal & OS & 376 & Med, Comb, Single & 1.51 & 0.49 & 5.4 \\
\hline
\end{tabular}
\tablefoot{The listed exposure time "Exp." is the shutter-open time for one object (O) or sky (S) series. The data files in parentheses are affected by instrumental problems (see Sect.~\ref{obs} and App.~\ref{GRAVITYplots}). The instrumental setup includes the spectral resolution (medium or high), the polarization (split or combined), and the field configuration (single or dual). The term "AM" refers to the airmass, $\sigma$ to the DIMM seeing, and $\tau_0$ to the coherence time.}
\end{table*}


\section{Analysis \label{analysis}}

\subsection{Predicted black hole properties}

\begin{figure}
     \centering
         \includegraphics[width=0.9\hsize]{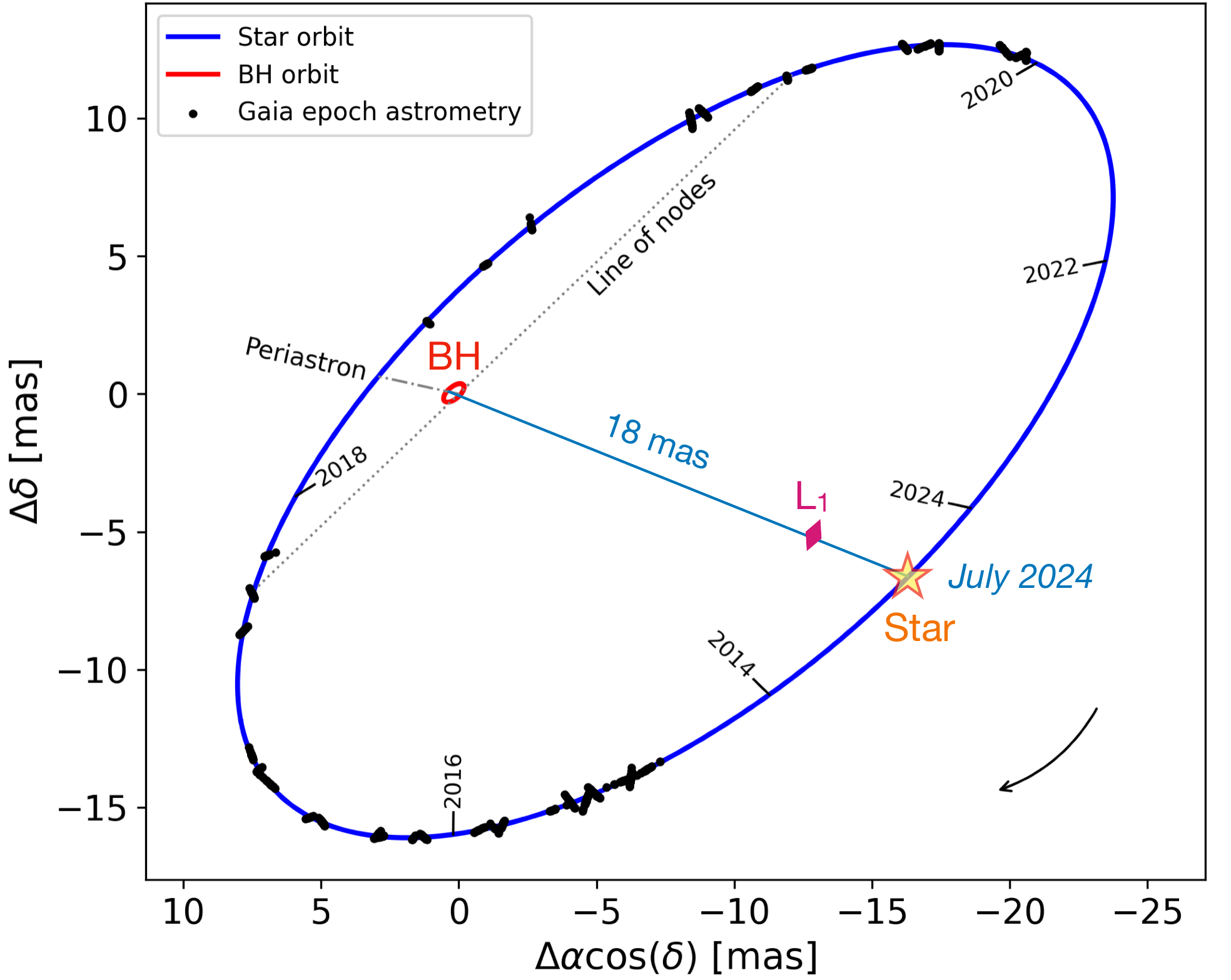}
     \caption{Position of Gaia BH3 and its companion star relative to their center of mass at the epoch of GRAVITY observations. The position of the Lagrangian L1 point of the system is shown with a purple diamond symbol.}\label{fig:bh3_orbit}
\end{figure}

The orbital elements determined from \textit{Gaia} astrometry and spectroscopic radial velocities by \citetads{2024A&A...686L...2G} provide a precise prediction of the positions of the BH and giant star components of Gaia BH3 (Fig.~\ref{fig:bh3_orbit}). On 23 June 2024, the positional offset from the giant star to the BH was $(\Delta\alpha,\Delta\delta)=(+16.7, +6.6)$\,mas, and it was $(+16.4, +6.9)$\,mas for the 18 July 2024 epoch.
Knowing the position facilitates the search for infrared emission at the position of the BH in the GRAVITY data.
The radial velocity of both the star and the BH is close to $v_R = -350$\,km\,s$^{-1}$ (the systemic radial velocity of Gaia BH3).
Gaia BH3 appears as a quiet BH in the UVES spectrum of the star, with no detectable emission lines \citepads{2024A&A...686L...2G}.
While the radius of the event horizon of the BH itself is very small ($\approx 96$\,km), its Roche lobe is very extended, thanks to its dominant mass.
The Lagrangian $L_1$ point of the system (i.e., the point between the two bodies where the gravitational forces from the star and the BH balance each other in the rotating frame) varies in position on the star-BH axis over the orbital cycle due to the eccentricity and varying orbital angular velocity. As shown in Fig.~\ref{fig:bh3_orbit}, the position of $L_1$ in the Gaia BH3 system is approximately at 20\% of the star-BH separation from the star toward the BH. 
Due to this peculiar geometrical configuration, a significant fraction of the wind emitted by the giant star is intercepted by the Roche lobe of the BH, and part of it may be accreted. The wind of the giant star may therefore ensure a steady flow toward the BH and a stable emission from accretion.

\subsection{Search for continuum emission \label{continuumsearch}}

\begin{figure}
     \centering
         \includegraphics[width=0.8\hsize]{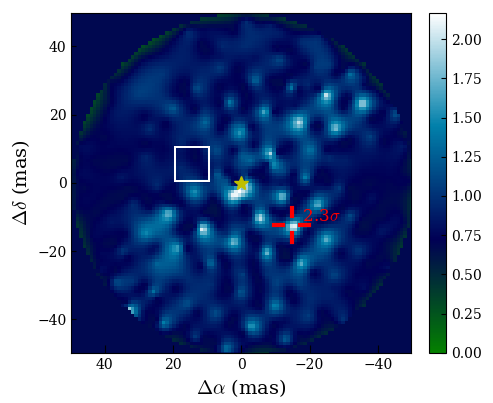}
         \includegraphics[width=0.55\hsize]{figures/carte_zoom.png}
     \caption{Top panel:  \texttt{CANDID} map of companion detection probability expressed in the number of standard deviations of the noise level for the combination of 23 June and 18 July 2024 data in medium spectral resolution. The S/N color scale is from 0 to $2.1\sigma$. The position of the star is marked with a star symbol, and the maximum S/N ratio of $2.3$ is indicated with a red cross. The predicted position of the BH is framed with a white square. Bottom panel: Contrast map for a 3$\sigma$ detection level around the expected position of the BH, where the color scale represents the contrast $\Delta m$ in magnitudes.}\label{fig:CANDID-map}
\end{figure}

We used the companion analysis tool \texttt{CANDID}\footnote{\texttt{https://ascl.net/1505.030}} \citepads{gallenne2015} to systematically search for the signature of Gaia BH3 in the GRAVITY data. \texttt{CANDID} uses a two-dimensional position grid and a multi-parameter fit based on a least-squares minimization algorithm.
We jointly considered the data from the 23 June and 18 July 2024 observing epochs in the search for BH emission, as the shift in position of the giant star is only $400\,\mu$as. We checked through a numerical simulation that this shift has a negligible impact on the sensitivity of the companion search.
The presence of a secondary emission in addition to the giant star in the field of view of GRAVITY has an influence on several interferometric observables and on the closure phase (CP; or T3PHI) in particular.
This observable has a low sensitivity to biases introduced by the atmospheric piston effect (time-dependent variation of the optical delay between the beams collected by each telescope). The CP is a very good tracer of any deviation from central symmetry of the light distribution in the field of view, which may be caused by a secondary source.
To ensure a sufficiently complete coverage of the $\chi^2$ map, we conducted the search on a grid of starting positions from where the $\chi^2$ of the fit is minimized.
The detection limit of \texttt{CANDID} is determined based on the injection and recovery of artificial companions over the search grid.
At each search position $(\Delta \alpha, \Delta \delta)$, \texttt{CANDID} outputs the relative flux with respect to the primary star of a secondary source compatible with the input data.
Examples of recent applications of this algorithm to the detection of faint companions are presented in \citetads{2024ApJ...972..145E} and \cite{2025A&A...693A.111G}. 
The \texttt{CANDID} maps of detection probability and sensitivity are shown in Fig.~\ref{fig:CANDID-map}.
At the predicted location of the BH, we determined a contrast limit of $\Delta m = 6.8$\,mag, that is, $C = f_\mathrm{BH}/f_\star = 0.19$\% over the $K$ band ($\lambda = 2.0-2.4\,\mu$m).
The expected flux of the giant star from its 2MASS $K_s$ magnitude \citepads{2006AJ....131.1163S} is
\begin{equation}
f_\mathrm{K, \star} = [1.01 \pm 0.01]\times 10^{-13}$\,W\,m$^{-2}$\,$\mu$m$^{-1}.
\end{equation}
The contrast limit translates into a BH flux upper limit of
\begin{equation}
f_\mathrm{K, BH} < 1.9 \times 10^{-16}$\,W\,m$^{-2}$\,$\mu$m$^{-1}.
\end{equation}

\begin{figure}
     \centering
         \includegraphics[width=\hsize]{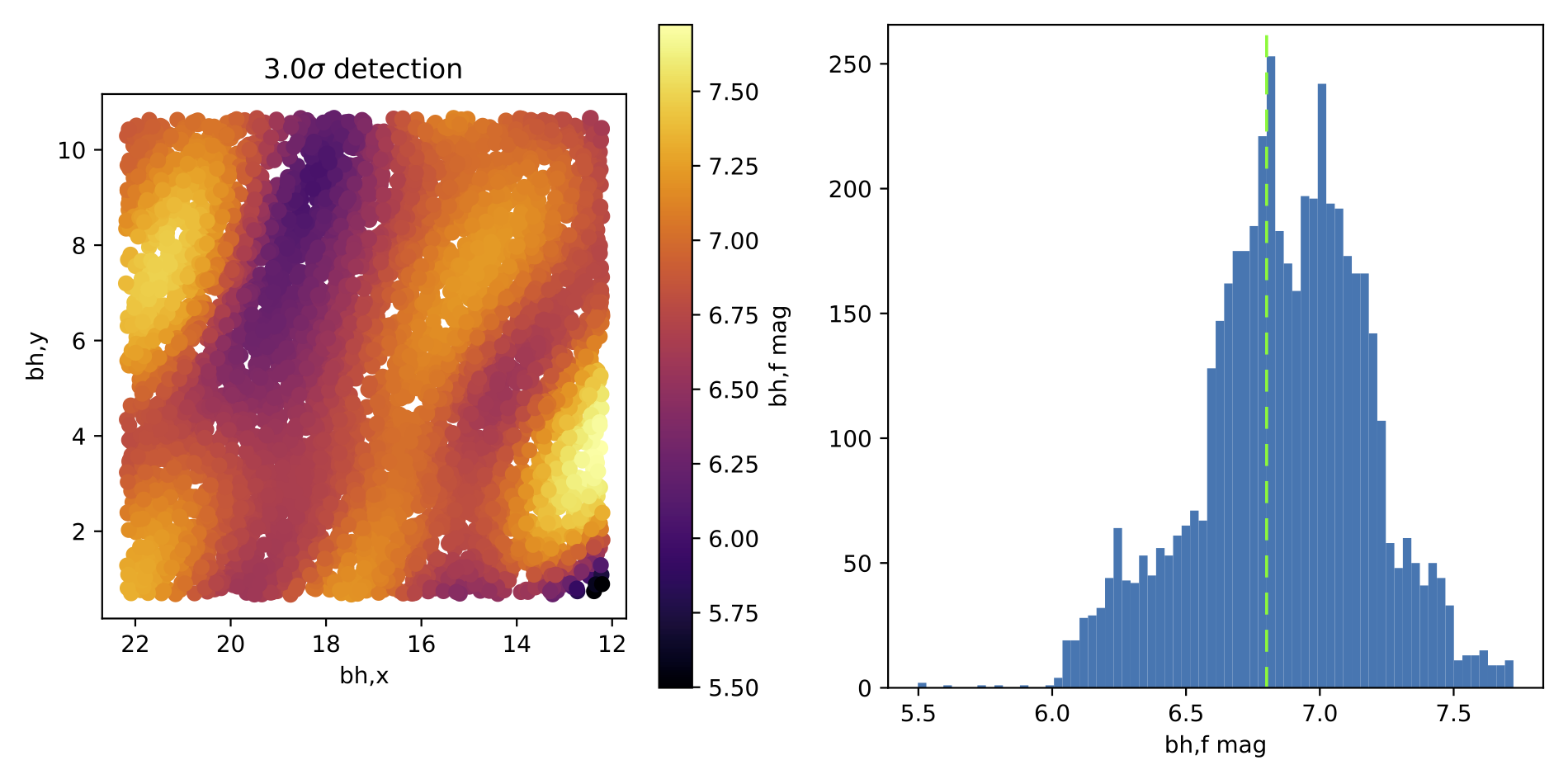}
     \caption{Left panel: \texttt{PMOIRED} map of the limiting magnitude ($3\sigma$, 99\% confidence) of the BH around its predicted position for the 23 June and 18 July 2024 combined data set. The displayed field corresponds to the white square in Fig.~\ref{fig:CANDID-map}.
     Right panel: Statistical distribution of the magnitudes of the recovered $3\sigma$ companions. Here, the dashed line represents the \texttt{CANDID} detection limit.}\label{fig:pmoired}
\end{figure}

To validate this \texttt{CANDID} detection limit, we also applied the \texttt{PMOIRED}\footnote{\url{https://ascl.net/2205.001}} code \citepads{2022SPIE12183E..1NM} to the combined 23 June and 18 July 2024 data set.
This library fits user-defined multi-component source models to model astronomical spectro-interferometric data.
The minimum $\chi^2$ is determined using a classical gradient descent method starting from a grid, resulting in best-fit values, uncertainties, and a correlation matrix. The contrast map around the position of the BH is presented in the left panel of Fig.~\ref{fig:pmoired}.
The simulated CPs (T3PHI) and normalized visibilities (N|V|) corresponding to an artificial companion with a flux of 5\% of that of the star at the predicted location of the BH are presented in Fig.~\ref{fig:PMOIREDsimu}.
These data are compared to the 23 June and 18 July 2024 combined data set to illustrate the absence of a significant detection of BH emission.
The \texttt{PMOIRED} limiting contrast for the continuum BH emission is consistent with the $\Delta m = 6.8$\,mag value obtained with \texttt{CANDID} (Fig.~\ref{fig:pmoired}, right panel).
Finally, we obtained a similar contrast limit compared to \texttt{CANDID} and \texttt{PMOIRED} by applying the \texttt{exoGravity}\footnote{\url{https://gitlab.obspm.fr/mnowak/exogravity}} code \citepads{2020A&A...633A.110G, 2020SPIE11446E..0OL}.

\subsection{Hydrogen Brackett $\gamma$ line \label{brGline}}

The high spectral resolution setting of the GRAVITY science combiner offers the possibility to search for BH emission in the hydrogen Br$\gamma$ line. At this wavelength, the contrast may be reduced due to the combination of absorption in the star's spectrum and emission around the BH. It is, however, less efficient in terms of contrast limit for the detection of continuum emission.
The 23 June 2024 high spectral resolution data do not exhibit any significant signature in CP and triple amplitude in the Br$\gamma$ line blue shifted at $2.164\,\mu$m ($v_R = -350$\,km\,s$^{-1}$; Fig.~\ref{fig:BrG1}).
We determined the detection limit in the Br$\gamma$ line from this data set by assuming the BH has a possible emission line with a velocity offset of $\pm30$\,km\,s$ ^{-1}$ around the expected velocity and a velocity dispersion $\delta v$ between 50 and 500\,km\,s$ ^{-1}$.
The maximum emission line strength for each given model was then estimated by computing the amplitude equivalent to a $3\sigma$ detection.
The 18 July 2024 medium resolution data also do not show any significant Br$\gamma$ signal (Fig.~\ref{fig:BrG2}).
As shown in Fig.~\ref{fig:BrGlimit}, the detection limit mostly depends on $\delta v$, and it reaches $\Delta m = 4.5$\,mag for $\delta v = 300$\,km\,s$^{-1}$.
This behavior can be intuitively explained as follows: The higher the velocity dispersion, the more spectral channels are involved in the potential detection.

\section{Conclusions \label{conclusion}}

\begin{figure}
     \centering
         \includegraphics[width=0.8\hsize]{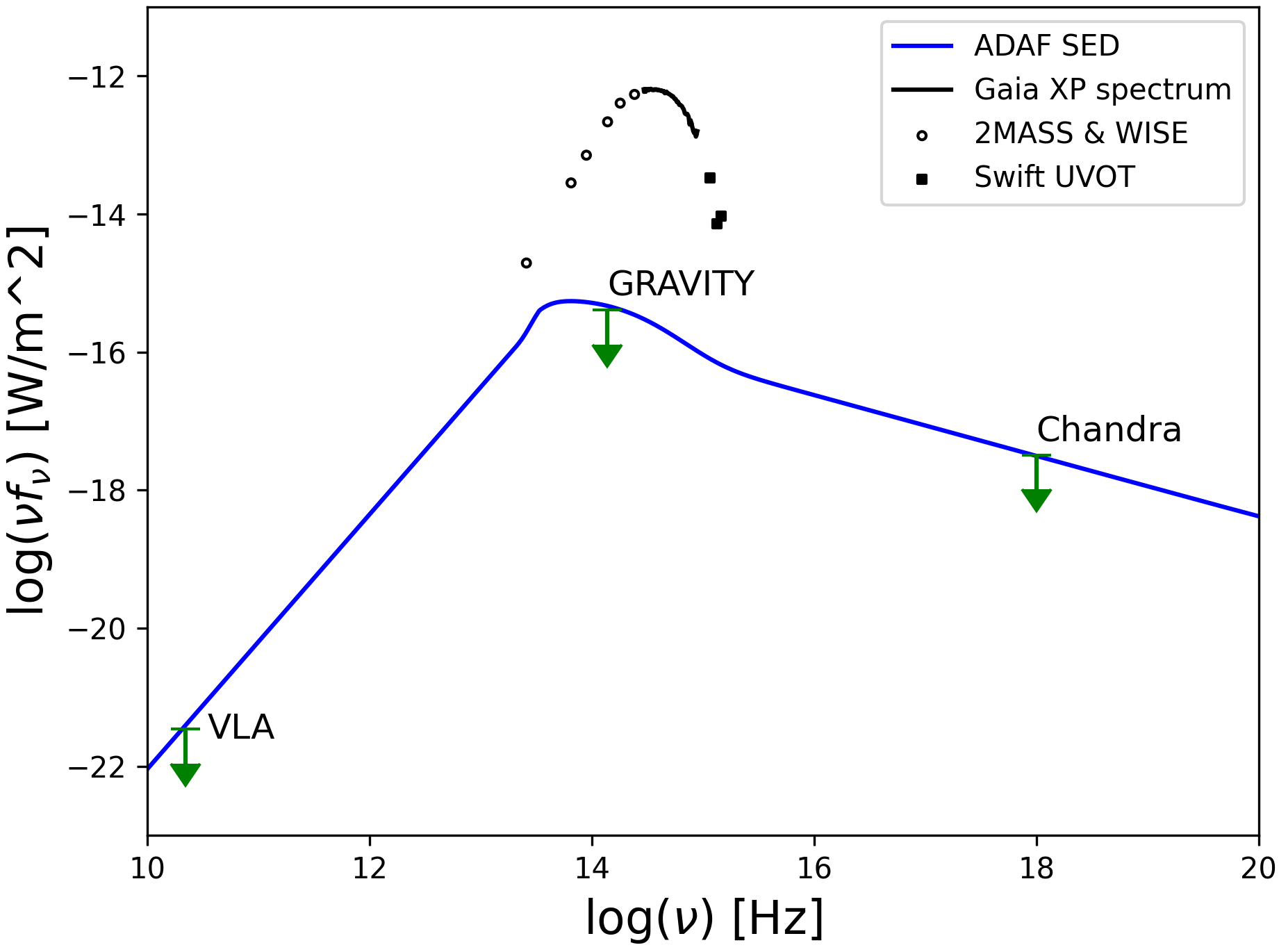}
     \caption{Upper limits on the spectral energy distribution of Gaia BH3 established from radio (August-September 2024; \citeads{2024ATel16832....1S, 2024ATel16677....1S}) and X-ray (May 2024; \citeads{2024ApJ...973...75C}) observations and the GRAVITY near-infrared observations (June-July 2024). The blue curve represents the accreting BH model presented by \citetads{2024ApJ...973...75C} for $f_\mathrm{Edd}< 4.9 \times 10^{-7}$.}\label{fig:model}
\end{figure}

We did not detect emission from the BH in Gaia BH3 in the $K$ band at a level of $f_\mathrm{K, BH3} < 1.9 \times 10^{-16}$\,W\,m$^{-2}$\,$\mu$m$^{-1}$.
This flux limit is shown in Fig.~\ref{fig:model} together with the upper limits established by \citetads{2024ApJ...973...75C} in the X-ray domain from Chandra ACIS-S observations and by \citetads{2024ATel16832....1S, 2024ATel16677....1S} in the radio domain ($\nu = 3, 10$ and 22\,GHz).
As discussed by \citetads{2024ApJ...973...75C}, accretion of the wind from the giant star (whose mass loss rate is estimated to be between $10^{-11}$ and $10^{-9}\,M_\odot\,a^{-1}$) by the BH is likely to occur through an advection-dominated accretion flow (ADAF; \citeads{Mahadevan_1997}).
In this configuration, the spectral energy distribution of the accreting BH including synchrotron, inverse Compton, and bremsstrahlung \citepads{Pesce_2021} is determined by the BH mass and the Eddington ratio $f_\mathrm{Edd}$.
This adimensional parameter is defined as the actual accretion rate divided by the Eddington rate (the accretion rate for which the BH radiates at the Eddington luminosity).
As shown in Fig.~\ref{fig:model}, the GRAVITY limit is consistent with the upper limit of $f_\mathrm{Edd}< 4.9 \times 10^{-7}$ derived by \citetads{2024ApJ...973...75C}.

Together with the non-detection in X-ray and radio domains of Gaia BH1 and BH2 recently reported by \citetads{2024PASP..136b4203R}, the nearest stellar mass BHs appear to be particularly quiet.
While the Gaia BH3 components are presently close to their maximum separation, the eccentricity of the orbit (Fig.~\ref{fig:bh3_orbit}) will bring them much closer together around 2030.
This may result in a significant increase of the accretion efficiency of the giant star wind by the BH and an up to $50$-fold boost in emission \citepads{2024ApJ...973...75C}.
The new GRAVITY+ Adaptive Optics (GPAO) system \citepads{2024arXiv240908438B, 2022Msngr.189...17A, 10.1117/12.3020042} will provide a threefold gain in photometric efficiency compared to the MACAO system for a bright target such as Gaia BH3.
The reported contrast limit on the BH emission relies on only three hours of effective GRAVITY observing time, and
deeper observations will realistically provide a gain in sensitivity by another factor of two. 
This combination of stronger emission at periastron and enhanced GRAVITY+ sensitivity by a factor of six opens up exciting perspectives to detect accretion-induced emission from Gaia BH3 toward the end of the 2020s.

\begin{acknowledgements}
We thank Drs Nico Cappellutti and Fabio Pacucci for sharing the BH emission model presented in Fig.~\ref{fig:model}.
Based on observations collected at the European Organisation for Astronomical Research in the Southern Hemisphere under ESO DDT programme 113.27R4.
We are grateful to ESO's Director General, Prof. Xavier Barcons, for the allocation of VLTI time to our program.
This work has made use of data from the European Space Agency (ESA) mission \textit{Gaia} (\url{http://www.cosmos.esa.int/gaia}), processed by the \textit{Gaia} Data Processing and Analysis Consortium (DPAC, \url{http://www.cosmos.esa.int/web/gaia/dpac/consortium}). Funding for the DPAC has been provided by national institutions, in particular the institutions participating in the {\it Gaia} Multilateral Agreement.
The research leading to these results has received funding from the European Research Council (ERC) under the European Union's Horizon 2020 research and innovation program (project UniverScale, grant agreement 951549).
MC acknowledges financial support from the Centre National d'Etudes Spatiales (CNES).
This research has made use of Astropy\footnote{Available at \url{http://www.astropy.org/}}, a community-developed core Python package for Astronomy \citepads{2013A&A...558A..33A,2018AJ....156..123A}, the Numpy library \citepads{numpy}, the Scipy library \citepads{scipy}, the Astroquery library \citepads{2019AJ....157...98G} and the Matplotlib graphics environment \citepads{Hunter:2007}.
We used the SIMBAD and VizieR databases and catalog access tool at the CDS, Strasbourg (France), and NASA's Astrophysics Data System Bibliographic Services. The original description of the VizieR service was published in \citetads{2000A&AS..143...23O}.
This publication makes use of data products from the Two Micron All Sky Survey, which is a joint project of the University of Massachusetts and the Infrared Processing and Analysis Center/California Institute of Technology, funded by the National Aeronautics and Space Administration and the National Science Foundation.
\end{acknowledgements}

\bibliographystyle{aa} 
\bibliography{Bibliography-Kervella}

{\onecolumn
\begin{appendix}

\section{GRAVITY data quality and contrast limits\label{GRAVITYplots}}

We here discuss the interferometric observables of the GRAVITY data collected on 3 June, 23 June, and 18 July 2024
For all epochs, the calibrator \object{BD+08 4132} was positioned on the optical axis of the fringe tracker (FT) and SC fibers, with the same spectral resolution and polarization setting as Gaia BH3.

\subsection{3 June 2024 \label{3june2024}}
The 3 June 2024 observations included a splitting of polarizations using a Wollaston prism inserted in the FT and SC beams \citepads{2024A&A...681A.115G}.
Both the SC and FT fibers were initially positioned on the giant star component of Gaia BH3 (single-field, on-axis mode).
Unfortunately, a first problem that we detected is that the input fiber of the FT and SC corresponding to the UT1 telescope drifted with respect to the target star during the observation. This resulted in a progressive loss of flux, and also in a displacement in phase of the fringes during the sequence. These symptoms are observed on all baselines that include UT1, and can be explained by an inaccurate sky tracking. We attribute the variation of the differential phase with wavelength to this drift. The drift in optical path difference (OPD) translates to a different phase offset depending on wavelength. Moreover, we see a differential behavior in polarization (blue and green curves in the U1-U3 panel of the top figure in Fig.~\ref{fig:dataplots030624}). This is likely also a symptom of the misalignment of the fibers on the target.
A second problem is that the visibility is increasing with wavelength in the upper panel of Fig.~\ref{fig:dataplots030624}, while the star being unresolved we expect a constant value of 1.0 (as in the bottom panel of Fig.~\ref{fig:dataplots230624med}). This indicates that the FT and SC channels were out of phase. The metrology was inactive during these observations, and although it is in principle not needed for single-field observations, this apparently caused an OPD offset between the FT (zero OPD) and the SC (non-zero OPD).
Considering the biases that are created by these instrumental problems, we did not include the data set of 3 June 2024 is our search for BH emission (however, we present the corresponding radial contrast limit in App.~\ref{contrastlimits}).

\begin{figure*}[h!]
     \centering
         \includegraphics[width=0.83\hsize]{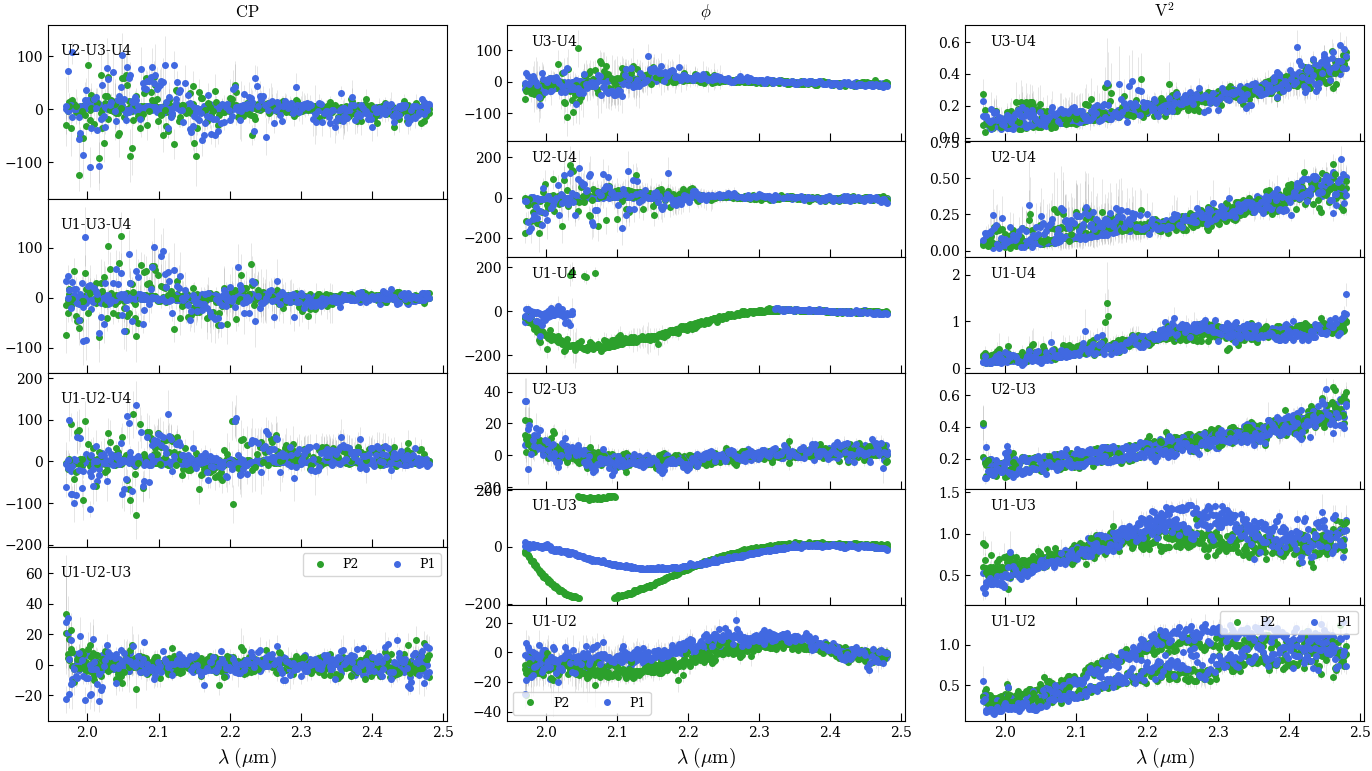}
     \caption{Interferometric observables collected on 3 June 2024 in medium spectral resolution with split polarizations (the blue and green points represent the two polarizations).
     In each panel, the three plots show the CPs (left), the spectral differential phases $\phi$ (center), and the squared visibilities $V^2$ (right) measured with GRAVITY on Gaia BH3.}\label{fig:dataplots030624}
\end{figure*}


\subsection{23 June 2024 \label{23june2024}}

On 23 June 2024, we tested both the medium and high spectral settings of the SC, respectively providing spectral resolutions of $R\approx 500$ and $4000$, while stabilizing the fringes with the FT. The two sets of fibers (FT and SC) were on-axis, that is, both positioned on the giant star.
As listed in Table~\ref{table:obslog}, the first 'object' exposure on Gaia BH3 in high spectral resolution mode starting at UT08:30:14 is affected by a large visibility bias and is discarded.

The feature present in Fig.~\ref{fig:dataplots230624med} around $2.16\,\mu$m is intriguing, as it could in principle indicate Br$\gamma$ line emission from the BH. However, this feature is unlikely to be of astrophysical origin for two reasons.
Firstly, no comparable signature is observed in CP. This is a consistency check as the differential phase amplitude reaches 5 to 10$^\circ$, and it should be detectable in CP if caused by BH emission.
Secondly, there is no apparent signal in the squared visibility $V^2$ at $2.16\,\mu$m, while a secondary source creating such a differential phase offset should induce a detectable signature.

The CP and $V^2$ signals that are used for companion search, however, behave correctly, with satisfactory quality control parameters. We therefore included the 23 June 2024 data in the estimate of the BH emission contrast limit in Sect.~\ref{continuumsearch}.

\begin{figure*}
     \centering
         \includegraphics[width=0.83\hsize]{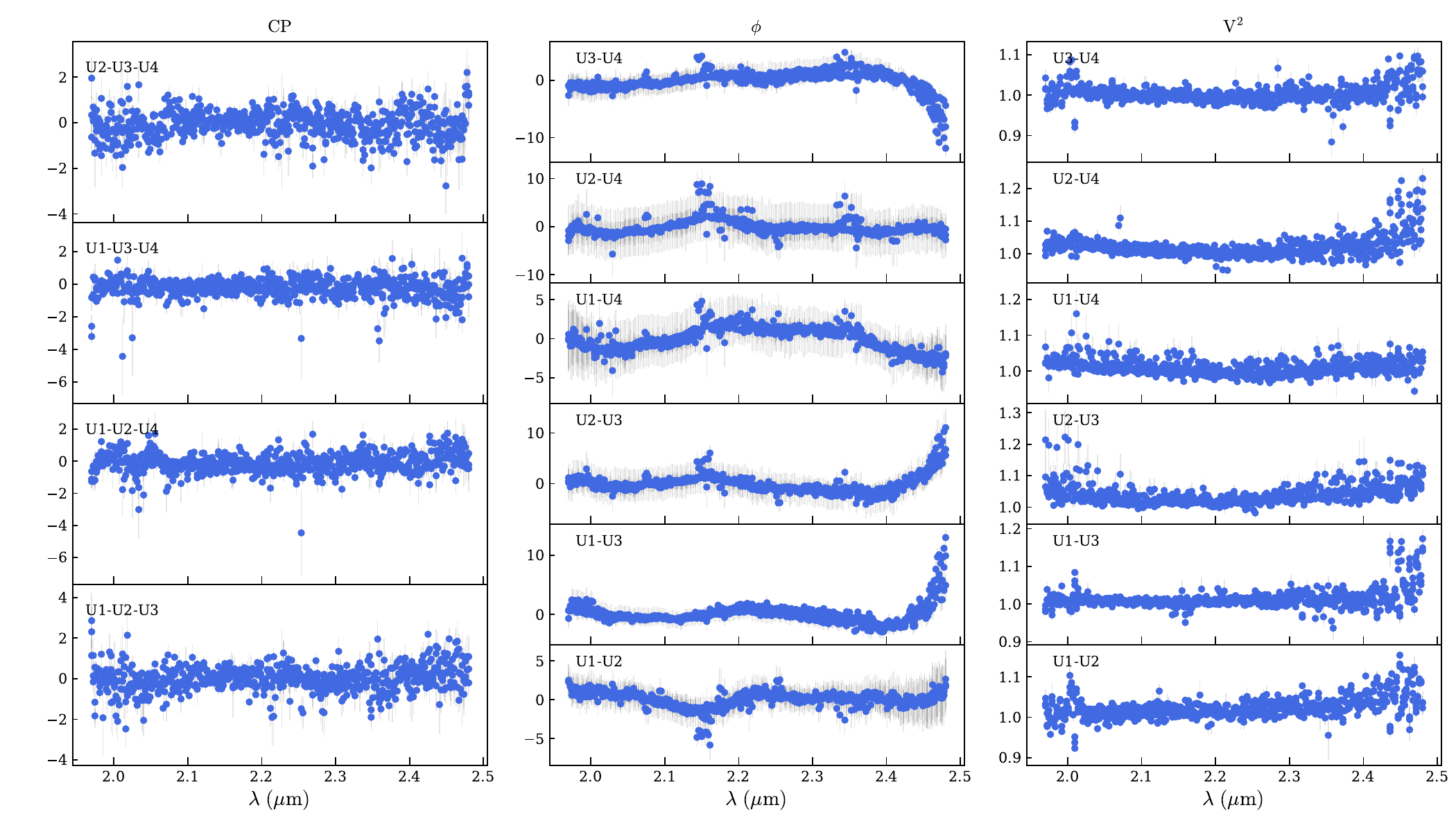}
     \caption{Same quantities as in Fig.~\ref{fig:dataplots030624} but for the observation of 23 June 2024 in medium spectral resolution with combined polarizations.}\label{fig:dataplots230624med}
\end{figure*}

\begin{figure*}
     \centering
         \includegraphics[width=0.83\hsize]{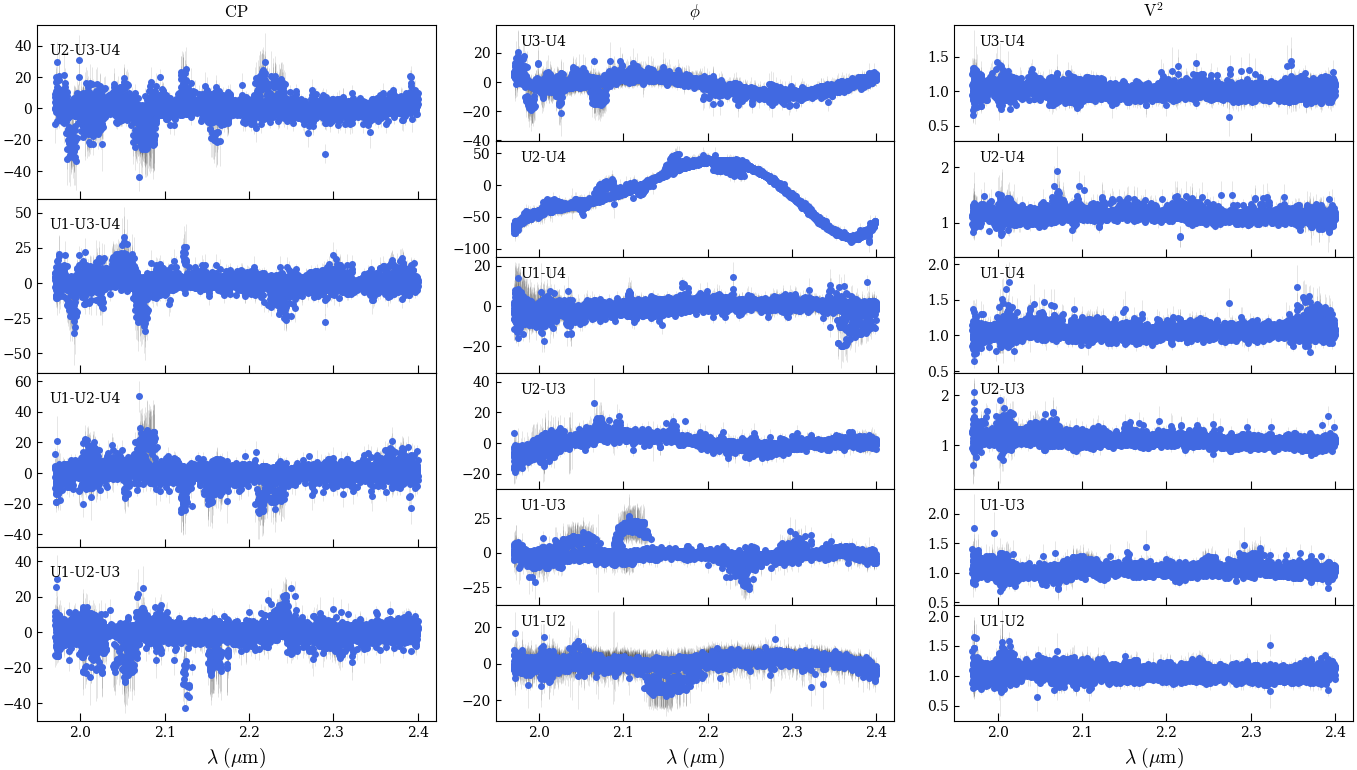}
     \caption{Same quantities as in Fig.~\ref{fig:dataplots030624} but for the observation of 23 June 2024 in high spectral resolution with combined polarizations.}\label{fig:dataplots230624high}
\end{figure*}


\subsection{18 July 2024 \label{18july2024}}

For our last observation, the SC fiber was offset from the FT fiber (dual-field, on-axis mode) by the expected star-BH relative position to optimize the injection of flux coming from the BH position in the SC. Due to the 18\,mas astrometric offset of the SC fibers with respect to the star's position, a wavelength-dependent phase shift of the SC fringes with respect to the FT fringes is visible in the central panel of the lower part of Fig.~\ref{fig:dataplots180724}. The large oscillations in differential phase $\phi$ are caused by phase folding. The $V^2$ and CP signals are consistent with the differential phases, and the photometric S/N level was high and stable during the whole observing sequence. The standard quality control parameters of this data set are all satisfactory.
The data obtained on 18 July 2024 is therefore included together with the 23 June 2024 epoch in our contrast limit determination presented in Sect.~\ref{continuumsearch}.

\begin{figure*}[h!]
     \centering
         \includegraphics[width=0.83\hsize]{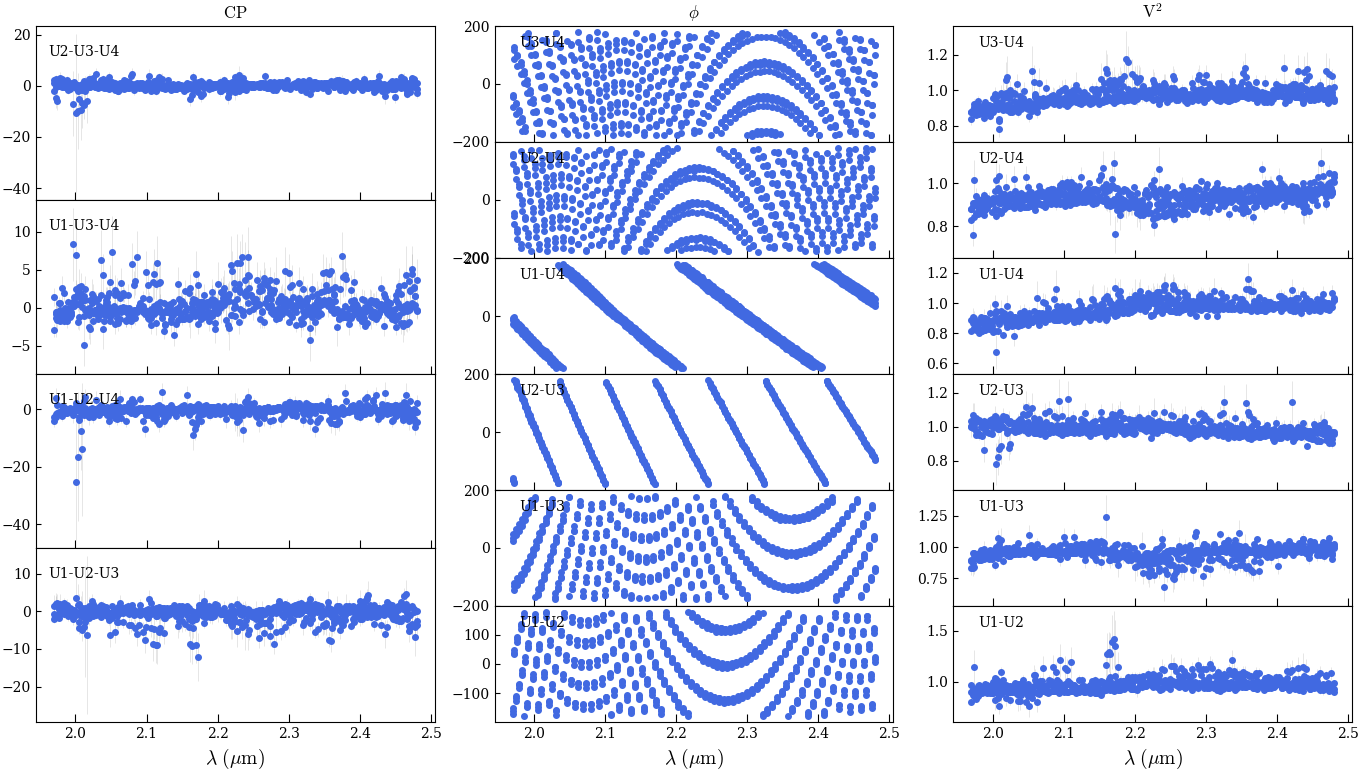}
     \caption{Interferometric observables collected on 18 July 2024 in medium spectral resolution with combined polarizations. This observation has been obtained with the SC fiber centered on the predicted BH position.
     The astrometric offset with respect to the star's position explains the large oscillations in differential phase $\phi$ as being due to phase folding.
     The plotted quantities are the same as in Fig.~\ref{fig:dataplots030624}.}\label{fig:dataplots180724}
\end{figure*}


\subsection{Contrast limits \label{contrastlimits}}

Fig.~\ref{fig:PMOIREDsimu} (left and central panels) shows a \texttt{PMOIRED} simulation of the interferometric CP (T3PHI) and normalized visibility (N|V|) signals that would be produced by a 5\% emission from the expected location of the BH (red curves). The normalized visibility is defined as $N|V| = V(\lambda)/L[V(\lambda)]$ where $L[V(\lambda)]$ is a linear fit to $V(\lambda)$ as a function of $\lambda$. The 18 July 2024 data set displayed as gray curves do not exhibit any signature of such a flux contribution.
The radial average dependence of the contrast limit established at each epoch from the medium spectral resolution observations is presented in the right panel of Fig.~\ref{fig:PMOIREDsimu}.
The improvement in sensitivity for the 23 June and 18 July 2024 compared to the 3 June 2024 epoch is clearly visible.
At the expected separation of the BH of 18\,mas in Fig.~\ref{fig:PMOIREDsimu} (right panel), the radial average contrast is slightly shallower than the $\delta m = 6.8$\,mag value reported in Sect.~\ref{continuumsearch}.
This is due to the fact that the particular $(u,v)$ plane coverage of the observations results in a deeper local sensitivity around the expected location of the BH (Fig.~\ref{fig:CANDID-map}) than its radial average.

\begin{figure*}[h!]
     \centering
         \includegraphics[height=5.6cm]{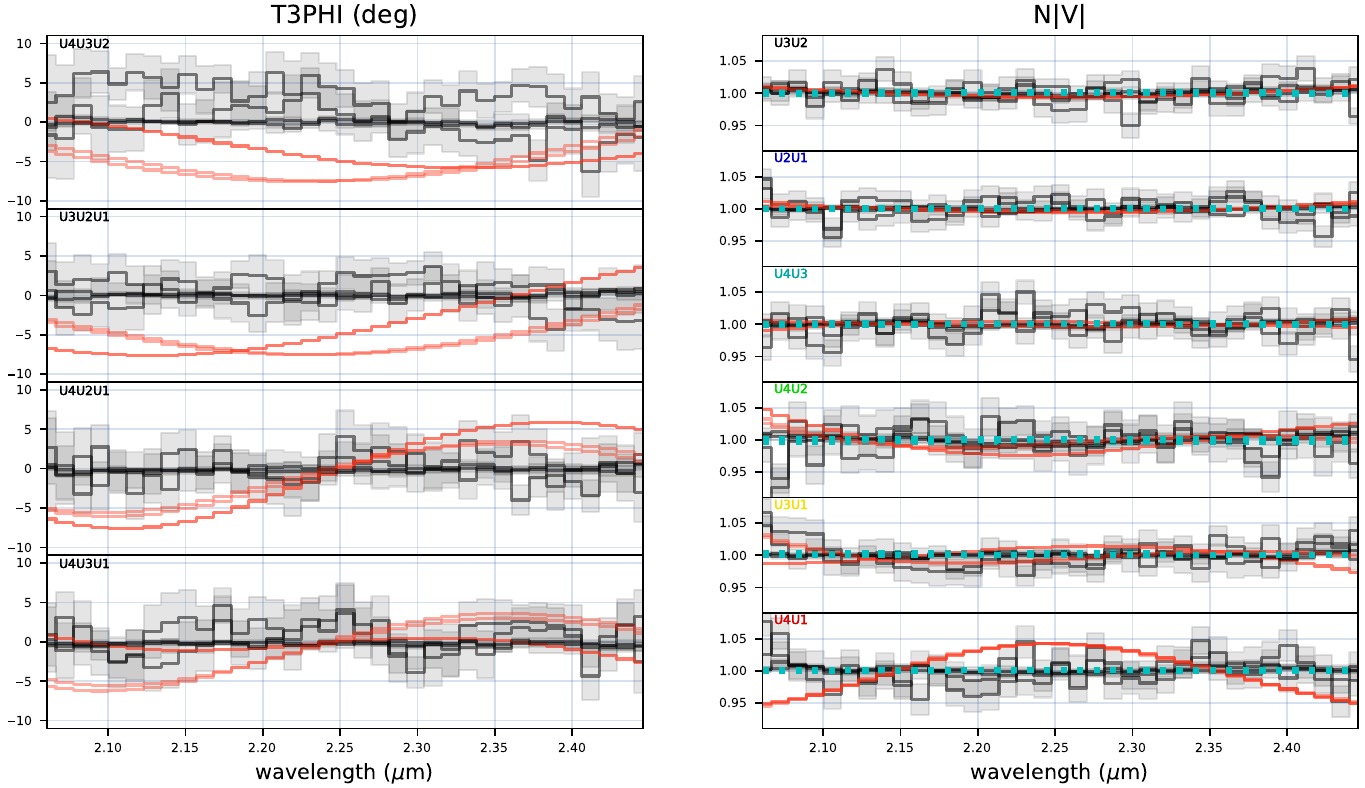}
         \includegraphics[height=5.6cm]{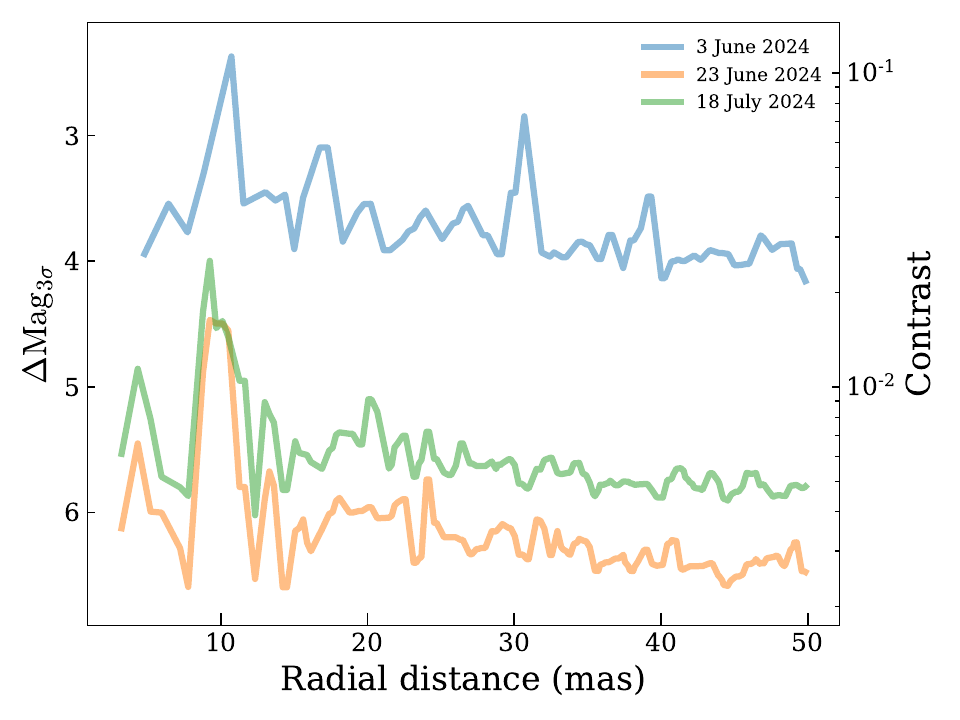}
     \caption{Left and central panels: Simulated CPs (T3PHI) and normalized visibility (N|V|) for a 5\% flux companion (red curves) injected in the data collected on 18 July 2024 (gray curves). Right panel: Contrast limits from \texttt{CANDID} as a function of the distance from the star for the three epochs in medium resolution. The contrast with 90\% confidence is shown in magnitude difference (left axis) and contrast (right axis).} \label{fig:PMOIREDsimu}
\end{figure*}


\subsection{High spectral resolution data and contrast limit\label{BrGplots}}

The CP and triple amplitude signals as a function of wavelength in the region of the Br$\gamma$ line are shown in Fig.~\ref{fig:BrG1} for the 23 June 2024 epoch (high and medium resolution) and in Fig.~\ref{fig:BrG2} for the 18 July 2024 epoch (medium resolution). 
The upper limit on the Br$\gamma$ emission from the BH based on the high spectral resolution data is shown in Fig.~\ref{fig:BrGlimit} and discussed in Sect.~\ref{brGline}.

\begin{figure*}
     \centering
         \includegraphics[width=0.73\hsize]{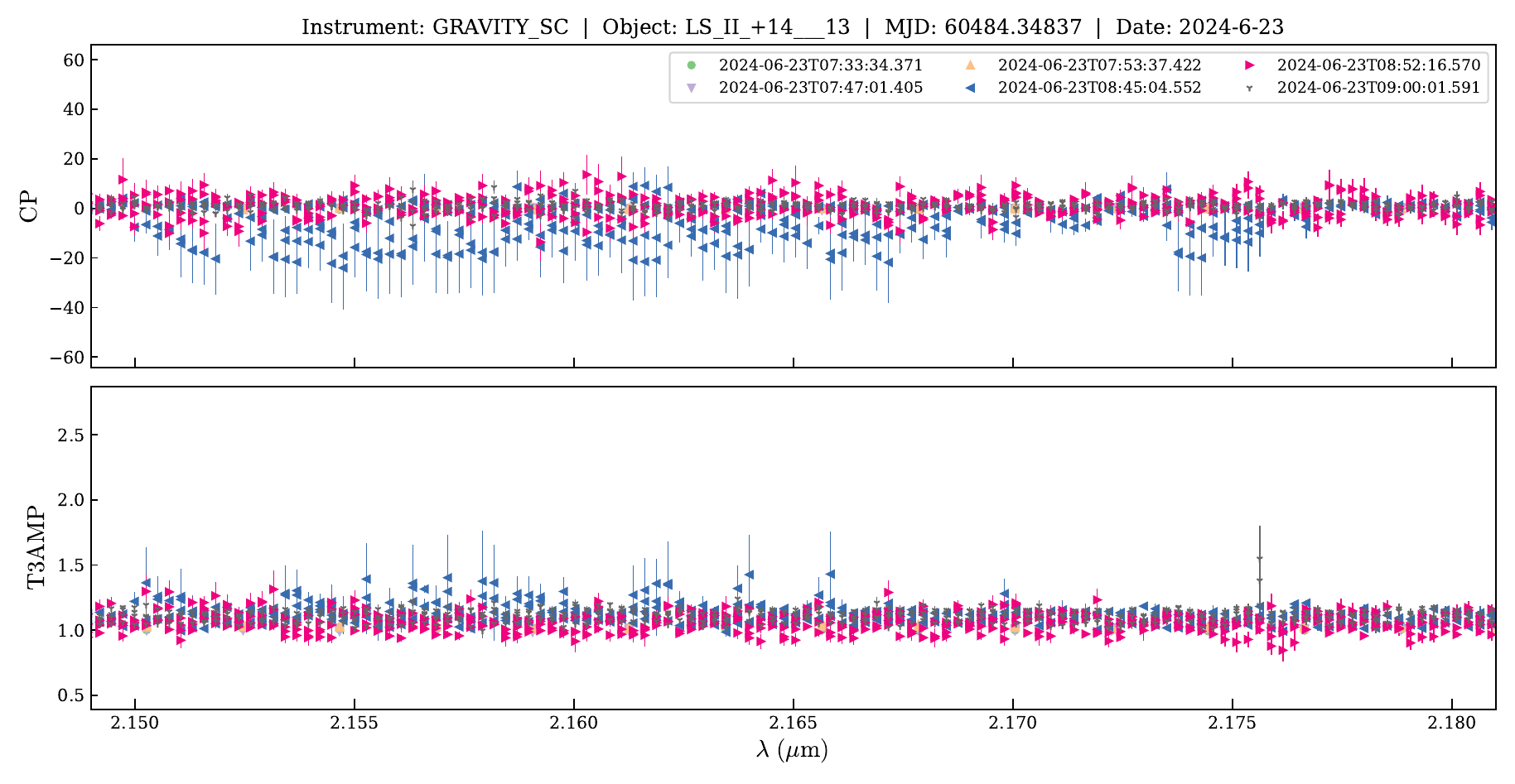}
     \caption{GRAVITY CPs (top panel) and triple amplitudes (bottom panel) of Gaia BH3 measured on 24 June 2024 (high and medium spectral resolutions) around the hydrogen Br$\gamma$ line (rest wavelength $\lambda=2.166\,\mu$m, Doppler-shifted in Gaia BH3 to $\lambda=2.164\,\mu$m).}\label{fig:BrG1}
\end{figure*}

\begin{figure*}
     \centering
         \includegraphics[width=0.73\hsize]{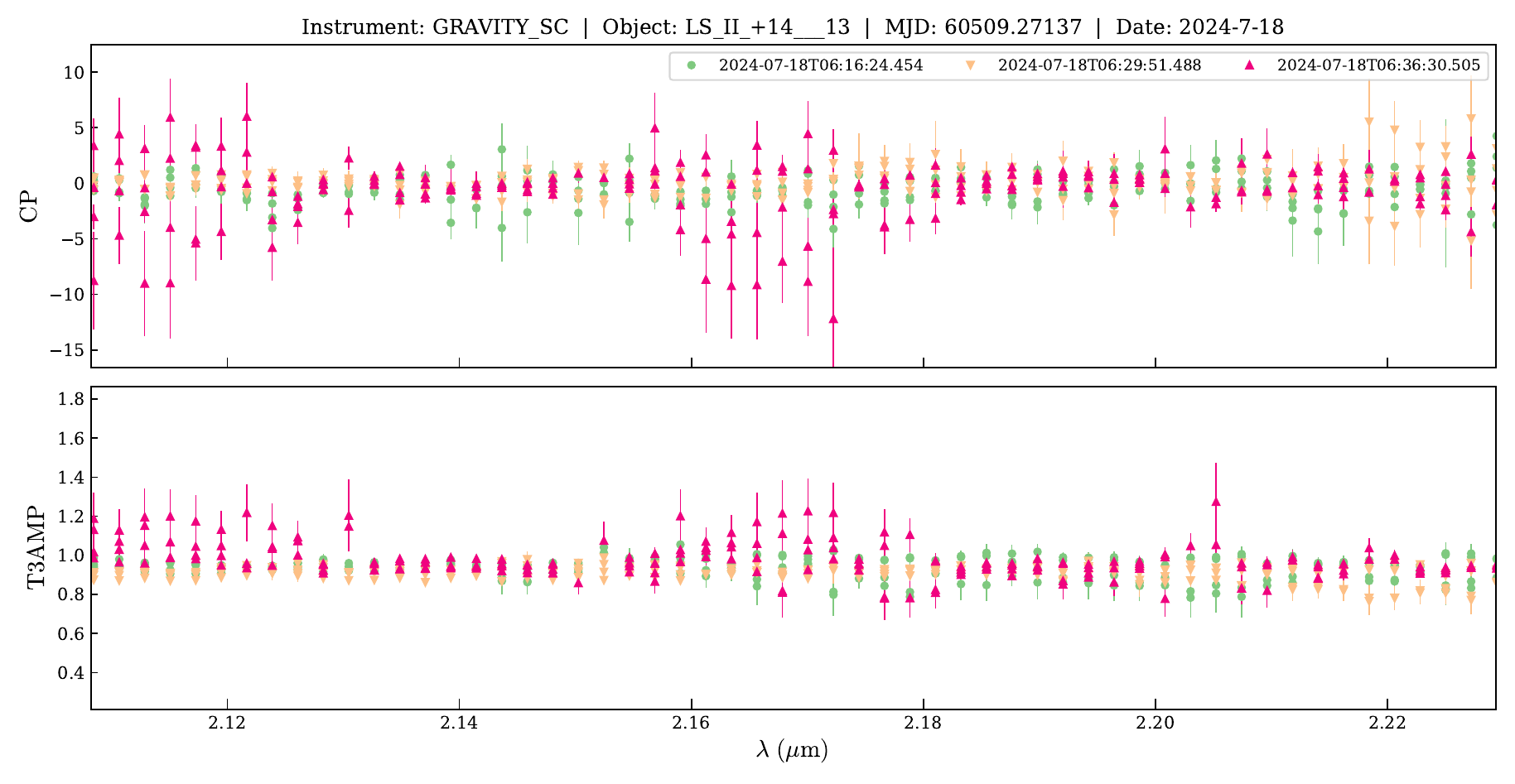}
     \caption{Closure phases (top panel) and triple amplitudes (bottom panel) of Gaia BH3 in the region of the hydrogen Br$\gamma$ line measured with GRAVITY on 18 July 2024 in medium spectral resolution mode with combined polarizations.}\label{fig:BrG2}
\end{figure*}

\begin{figure}
     \centering
         \includegraphics[width=0.47\hsize]{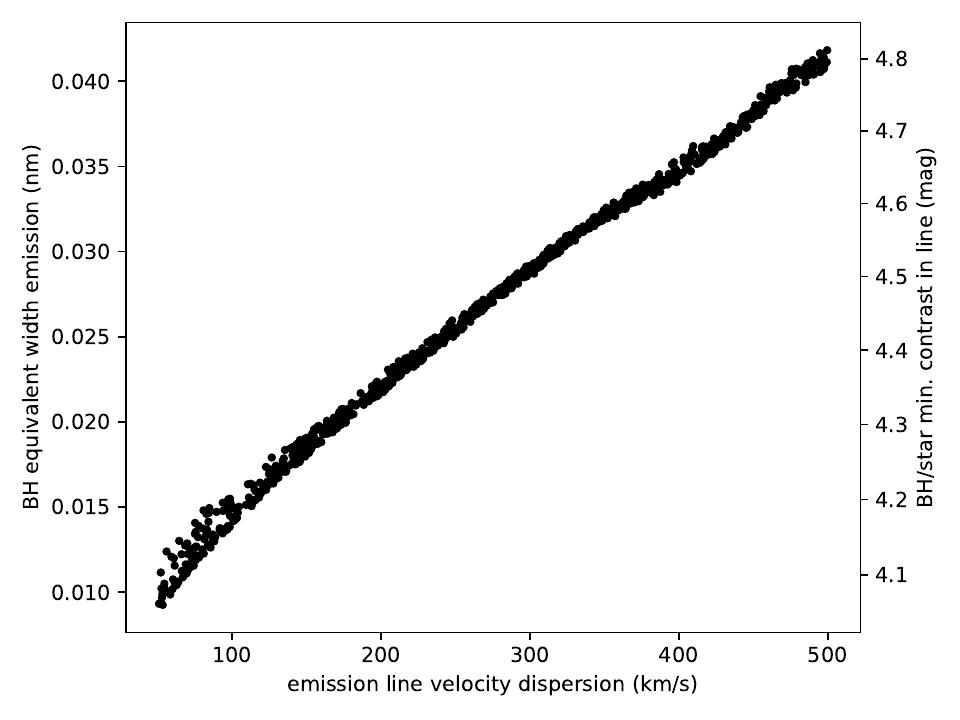}
     \caption{Upper limit of the equivalent width (left scale) and contrast limit (right scale) in the Brackett $\gamma$ line emission at the astrometric position of the BH component of Gaia BH3 as a function of the assumed emission line velocity dispersion.}\label{fig:BrGlimit}
\end{figure}

\end{appendix}
}

\end{document}